 \documentclass[aps,prx,notitlepage,superscriptaddress,twocolumn]{revtex4}
\usepackage{graphicx,subfigure,epsfig}
\usepackage{dcolumn}
\usepackage{amssymb}
\usepackage{times}
\usepackage{amsmath}
\usepackage{amsfonts}
\usepackage{mathrsfs}
\usepackage{setspace}
\usepackage{latexsym}
\usepackage{bbm}
\usepackage{float}
\usepackage{flafter}
\usepackage{bm}
\usepackage{epstopdf}
\usepackage{hyperref}
\usepackage{color}
\usepackage{multirow}
\usepackage{amsmath,amssymb}
\usepackage[flushleft]{threeparttable}
\usepackage{array,booktabs,makecell}
\usepackage[font=small]{caption}

\def \be {\begin{eqnarray}}
\def \ee {\end{eqnarray}}

\begin{document}

\title{Pseudoscalar U(1) spin liquids in $\alpha$-RuCl$_3$}

\author{Inti Sodemann Villadiego}
\affiliation{Max-Planck Institute for the Physics of Complex Systems, D-01187 Dresden, Germany}
\affiliation{Department of Physics and Astronomy, University of California, Irvine, California 92697, USA }

\begin{abstract}
A recent experiment has reported oscillations of the thermal conductivity of $\alpha$-RuCl$_3$ driven by an in-plane magnetic field that are reminiscent of the quantum oscillations in metals. At first glance, these observations are consistent with the presence of the long-sought-after spinon Fermi surface state. Strikingly, however, the experiment also reported vanishing thermal Hall conductivity coexisting with the oscillations of the longitudinal one. Such absence of the thermal Hall effect must originate from crystalline symmetries of $\alpha$-RuCl$_3$. But if the system was a traditional spinon fermi surface state, these symmetries would also necessarily prohibit the emergence of a magnetic field acting on the spinons, in stark contradiction with the presence of quantum oscillations in experiments. To reconcile these observations, we introduce a new class of symmetry enriched ``pseudoscalar" U(1) spin liquids in which certain crystalline symmetries act as a particle-hole conjugation on the spinons. The associated pseudoscalar spinon Fermi surface states allow for the coexistence of an emergent Landau quantizing magnetic field while having an exactly zero thermal Hall conductivity. We develop a general theory of these states by constructing Gutzwiller-projected wave-functions and describing how they naturally appear as U(1) spin liquids with a distinctive projective symmetry group implementation of crystalline symmetries in the fermionic parton representation of spins. We propose that the field induced quantum disordered state in $\alpha$-RuCl$_3$ descends from a pseudoscalar spinon fermi surface state that features compensated spinon-particle and spinon-hole pockets possibly located around the $M$ points of its honeycomb Brillouin zone. These points are connected via a wave-vector associated with the emergence of the competing zig-zag antiferromagnetic state.
\end{abstract}

\maketitle

\begin{figure}[!t]
\begin{center}
\includegraphics[width=0.47\textwidth]{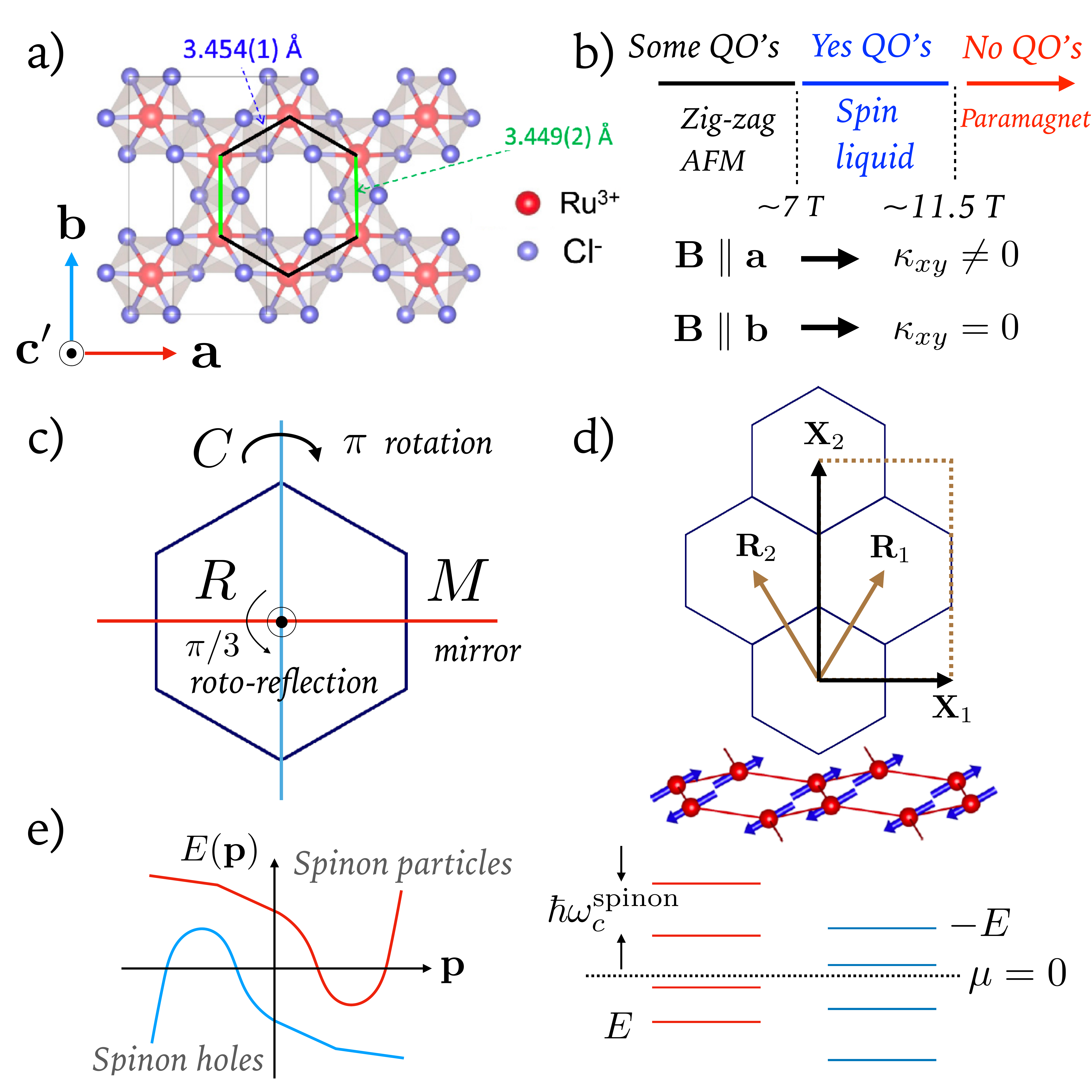}
\end{center}
\par
\caption{(Color online) (a) $\alpha$-RuCl$_3$ lattice and axis convention. (b) Summary of phase diagram and observations in Ref.~\cite{OngQO}. (c) Point symmetries of the contracted (mirror $M$ and $\pi$ rotation $C$) and un-contracted ($\pi/3$ roto-reflection $R$) $\alpha$-RuCl$_3$ lattice. (d) Bravais lattice vectors (${\bf R_{1,2}}$) and magnetic Bravais vectors (${\bf X_{1,2}}$) for the zig-zag AFM state depicted in lower inset. (e) pseudoscalar spinon Fermi surface particle- and hole-like pockets and the corresponding spinon Landau level spectrum resulting from an emergent magnetic field. Figure (a) and lower inset (d) were taken from Ref.\cite{Nagler_2016}.}
\label{Fig:abc}
\end{figure}

\section{Introduction}
\label{intro}

 A recent remarkable experiment~\cite{OngQO} has detected oscillations of the thermal conductivity of $\alpha$-RuCl$_3$ induced by an in-plane magnetic field. Some of its key observations, that will serve as cornerstones for our theoretical construction, are:

 {\bf Observation 1}. Quantum oscillations the longitudinal thermal Hall conductivity are induced by fields either along the ${\bf a}$ or the ${\bf b}$ axis, but their periods are largely independent of the out-of-plane component along the ${\bf c}'$ axis (see Fig.~\Ref{Fig:abc}).
 
{\bf Observation 2}. Quantum oscillations separate into two regimes, one with smaller amplitudes and frequency, in the field region of $B\sim 4T-7T$, that would coexist with the zig-zag AFM state~\cite{Coldea_2015,Nagler_2016,Nagler_2018,Nagler_2019}. And another with larger amplitudes and frequencies for  fields $B\sim 7T-11.5T$. 


{\bf Observation 3}. Thermal Hall effect is seen for fields along the ${\bf a}$ axis, in agreement with Refs.~\cite{Matsuda_2018,Matsuda_2018B,Hess_2019,Matsuda_2020,Takagi_2021}, but it is essentially absent for fields along the ${\bf b}$ axis, in agreement with Ref.~\cite{Matsuda_2020}. When present, the thermal Hall conductivity is much smaller than the oscillatory part of the longitudinal one (by about four orders of magnitude at $T\sim1K$), and it decreases as the temperature is lowered in a way that anti-correlates with the relative growth of the oscillatory longitudinal conductivity.

 
 
 
 
 
The above observations provide a strong motivation to conjecture the existence of a field induced quantum spin liquid featuring a Fermi surface of neutral Fermions~\cite{OngQO}, which have also been argued to be present in numerical studies of the Heisenberg-Kitaev models of $\alpha$-RuCl$_3$~\cite{Trivedi,Hickey,Jiang,Lu}. The oscillations also indicate that these fermions experience an emergent magnetic field different from the externally applied one, and therefore, that there is an emergent U(1) gauge field coupled to these fermions. Since $\alpha$-RuCl$_3$ is a layered Van-der-Waals material it is also natural to expect that much of the essential physics occurs within a single $\alpha$-RuCl$_3$ layer, and therefore, that the state is essentially 2D in nature. 

At first glance, the aforementioned characteristics are present in a phase of matter that has been a holy grail of condensed matter research since the pioneering work of Anderson and collaborators~\cite{ANDERSONScience,Baskaran-Anderson1,Baskaran-Anderson2}, known as the spinon Fermi surface state (see~\cite{KanodaRMP,Savary,Broholm} for reviews). The traditional version of this state can be understood from a slave boson (or the closely related slave rotor~\cite{rotors}) parton decomposition of the electron operator ($c^\dagger$) into a product of a spinful fermion (spinon $f^\dagger$) and a spinless boson (chargon $b^\dagger$):

\be\label{slaveboson}
c^\dagger_{{\bf r}s} = f^\dagger_{{\bf r}s} b^\dagger_{\bf r}.
\ee  

Here ${\bf r}$ is the lattice site and $s\in \{\uparrow,\downarrow\}$ the spin. The spinon Fermi surface state appears when the chargons form a trivial bosonic Mott insulator, while the spinons form a Fermi surface state. In spite of being an insulator to DC charge transport, the spinon Fermi surface state can display quantum oscillations in response to applied magnetic fields~\cite{Motrunich,Debanjan,SodemannPRB} (see Refs.~\cite{PatrickPRB,RaoPRB,Khoo_arxiv} for other amusing properties).  One could therefore be tempted to conjecture that $\alpha$-RuCl$_3$ harbours a spinon Fermi surface state, albeit with a rather peculiar coupling between the applied and emergent magnetic fields, such that a purely in-plane applied magnetic field in the ${\bf a}$-${\bf b}$ plane, induces a large orbital component of emergent magnetic field along the ${\bf c'}$ direction.

There is, however, a striking observation that strongly clashes with the traditional spinon Fermi surface scenario. Namely {\bf Observation 3}, according to which the system displays no thermal Hall effect when the external field is along the ${\bf b}$ axis and a negligible one when it is along the ${\bf a}$ axis (see Fig.~\Ref{Fig:abc} for the axis convention). In fact, at small fields, one expects that the spinons would contribute to the thermal Hall conductivity with a typical value $\kappa_{xy} \sim (\omega_c \tau) \kappa_{xx}$~\cite{Debanjan}, where $\omega_c$ is their effective cyclotron energy scale associated with the emergent magnetic field. However, the visibility of quantum oscillations requires that $\omega_c \tau \gtrsim 1$, and therefore, one typically expects a comparable contribution of the spinons to the oscillatory component of both $\kappa_{xx}$ and $\kappa_{xy}$ in the regime in which these are experimentally detectable. 

Such clear absence of thermal Hall effect when the field is along ${\bf b}$ must be the result of symmetries of $\alpha$-RuCl$_3$ that remain unbroken throughout the range of in-plane fields that includes both the zig-zag AFM and the quantum spin liquid state. There are two different symmetries of the underlying Hamiltonian of $\alpha$-RuCl$_3$ that are present when the field is along ${\bf b}$ (but are broken when the field is along ${\bf a}$), that would forbid the thermal Hall effect. Namely, the mirror that sends ${\bf b}\rightarrow -{\bf b}$, and a $\pi$ rotation around ${\bf b}$, that sends ${\bf a}\rightarrow -{\bf a}$ (in the AFM this symmetry survives in combination with a half-translation see Fig.~\Ref{Fig:abc}). However, in the traditional spinon Fermi surface scenario (Eq.~\eqref{slaveboson}), the emergent orbital magnetic field is odd under these symmetries, and therefore it would be altogether absent if these symmetries were present, in contradiction with the very presence of quantum oscillations when the external field is applied along the ${\bf b}$ axis.

The above leads us to conjecture that the emergent magnetic field experienced by spinons in $\alpha$-RuCl$_3$ is not a pseudoscalar (or equivalently and out-of-plane pseudo-vector along the ${\bf c'}$ axis), as it is the case of the familiar magnetic field experienced by electrons or the emergent magnetic field of the traditional spinon Fermi surface scenario, but rather that it is a {\it scalar} (even) under the aforementioned mirror and $\pi$-rotation symmetries. In order to naturally combine such transformation law with an emergent U(1) gauge structure, we will postulate that the spinon particle number is a pseudoscalar (odd) under these symmetries, or, in other words, that these symmetries act on the spinon as a particle-hole conjugation. Because of this property, we will refer to these states as ``psedoscalar U(1) spin liquids". 

But how could the spinons in such ``pseudoscalar spinon Fermi surface state" experience an emergent orbital magnetic field with its associated Lorentz force, while at the same have no net thermal Hall conductivity?. As we will show, these states naturally feature pairs of spinon particle-like and hole-like pockets of equal size related by point group symmetries. Therefore there would be an amount of particle-like spinons executing say clockwise cyclotron motion that is exactly balanced by the same amount of hole-like spinons executing the opposite counter-clockwise cyclotron motion, leading to an exact vanishing of the Hall conductivity enforced by the symmetry in question (see Fig.~\Ref{Fig:abc}(c)). This, however, does not preclude the existence of oscillations of the spinon density of states near the chemical potential, and hence the presence of quantum oscillations. 

We will not provide a detailed justification for why these states are energetically favorable in $\alpha$-RuCl$_3$ starting from a specific microscopic Hamiltonian. This is clearly a very important question, but at the moment we believe it is a conservative standpoint to remain partly agnostic on this issue given the uncertainties surrounding the ideal microscopic of model $\alpha$-RuCl$_3$~\cite{Chernyshev}, not to mention the challenges of solving such ideal models that feature large deviations from the ideal Kitaev model~\cite{Kitaev} and the relatively poorly understood emergence of even the traditional spinon Fermi surface state from ideal microscopic models (see however Refs.~\cite{Motrunich2,SSLee,DNSheng,Lai,Block,Mishmash} for analytical and numerical studies). We will however make an effort to provide not only an effective low-energy field theory description but also to understand microscopic aspects of the lattice scale anatomy of these states, because we view these experiments as an unprecedented opportunity to deepen our understanding of the emergence of U(1) spin liquids featuring gapless neutral fermions from real materials and also from microscopic models.

We have organized our paper by introducing first some of the most general ideas in Sec.~\Ref{Pseudo-general}. This section can be read indepently from the rest of the paper and does not rely on details of $\alpha$-RuCl$_3$. This section introduces the notion of pseudoscalar U(1) spin liquids and describes some of its properties in a general setting that could be relevant to other materials and models. The subsequent sections, however, rely heavily on the ideas from Sec.~\Ref{Pseudo-general}. Specificallly Sec.~\Ref{pseudoRuCl} deals with the symmetries of $\alpha$-RuCl$_3$ and discusses a concrete possible pseudoscalar state that features Fermi pockets at the M points which are connected by the wave-vector that naturally explains the emergence of the zig-zag state via a spinon particle-hole pair condensation. Sec.~\Ref{lowEmodel} introduces a simplified low energy model that makes calculations more amenable, and also provides certain rationale for why the M points are natural locations for the spinon pockets. This section provides explicit calculations for the quantum oscillations of equilibrium thermodynamic quantities like the magnetization. Finally Sec.~\ref{discussion} discusses the relation of the pseudoscalar U(1) spin liquids to ideal models and theories, and also describes suggestions for future experiments in $\alpha$-RuCl$_3$ and the possible relevance of the pseudoscalar U(1) spin liquids to other material candidates.


\section{pseudoscalar U(1) spin liquids}
\label{Pseudo-general}


To formulate our ideas we will imagine that the physics of interest can be described within a Hilbert space of spin-1/2 degrees of freedom residing at lattice sites labeled by ${\bf r}$. We will also imagine that the system is two-dimensional. We will employ the fermionic parton representation of the spins~\cite{Auerbach}, by writing the spin operators at each site as follows:

\be\label{sigmaalpha}
\sigma^\alpha_{{\bf r}}= {\bf \sigma}^\alpha_{ss'} f^\dagger_{{{\bf r}}s}  f_{{{\bf r}}s'},
\ee


where ${\bf \sigma}^\alpha_{ss'}$ are the elements of the $\alpha$ Pauli matrix. The fermion operator $f^\dagger_{{{\bf r}}s}$ creates a spinon with spin $s$ at site ${\bf r}$. Only the subspace with a single spinon at each site is physical, and therefore given a state of spinons $|\Psi_0 \rangle$, the physical state $|\Psi \rangle$, can be obtained via Gutzwiller projection as follows:

\be\label{Gutz}
|\Psi \rangle= \prod_{{\bf r}} \left(\frac{1-(-1)^{\sum_{s}f^\dagger_{{\bf r} s} f_{{\bf r} s}}}{2}\right) |\Psi_0 \rangle.
 \ee
 

The Gutzwiller projection is a non-trivial operation that generically impedes exact calculation of expectation values of the Hamiltonian or other physically relevant operators. It is possible, however, to develop a precise understanding of the symmetry properties of the Gutzwiller projected physical state, and this will be the working horse of our study. The action of symmetry in fractionalized phases can be quite rich, and includes possibilities beyond the black-and-white distinction of symmetric versus symmetry breaking of traditional phases, such as the fractionalization of quantum numbers via projective symmetry implementations~\cite{Wen_PSG1,Wen_PSG2} and even more subtle patterns such as ``weak-symmetry breaking"~\cite{Wen_2003,Kitaev,Maissam_2019,Rao_2021}. 

To illustrate the possible non-trivial symmetry implementations in our context, let us consider the action of a spatial mirror symmetry $M$ normal to the y-axis, namely one acting on the physical spins as follows:



\be
M \sigma^y_{\bf r} M^{-1}= \sigma^y_{M{\bf r}}, M \sigma^{x,z}_{\bf r} M^{-1}=- \sigma^{x,z}_{M{\bf r}},
\ee


where ${M}{\bf r}$ is the image of the site ${\bf r}$ under $M$. One possible representation of the action of this symmetry on the spinons, that we denote as $M_{\bf +}$, is that it acts as it would on ordinary electrons:

\be\label{Me}
M_{\bf +} f^\dagger_{{{\bf r}}s} (M_{\bf +})^{-1}= e^{i\vartheta^{+}_{{\bf r}}}  i\sigma^{y}_{ss'} f^\dagger_{{M}{\bf r}s'}.
\ee

where the factor $e^{i\vartheta^{+}_{{\bf r}}}$ accounts for possible extra phases in the implementation. There is, however, another possible representation of this same physical symmetry, that we denote as $M_{\bf-}$, which acts on the spinons as a particle-hole conjugation:

\be\label{Mm}
M_{\bf-} f^\dagger_{{{\bf r}}s} (M_{\bf-})^{-1}= e^{i\vartheta^-_{{\bf r}}} f_{{M}{\bf r} s}.
\ee

Where $e^{i\vartheta^-_{{\bf r}}}$ accounts for possible extra phases. One can readily verify that that these two operations lead to an {\it identical} action on all the physical spin operators defined in Eq.~\eqref{sigmaalpha}. Because both representations lead to the same transformation of the physical spins, and because all the spin operators are left invariant under two consecutive actions of $M$, it follows that the product $M_{\bf+} M_{\bf-}$ leaves all the physical spins invariant. The group containing all of such operations that act non-trivially on the fermions but that leave all the spin operators invariant is called the the parton gauge group. These operations act trivially within the physical sub-space and therefore are not physical symmetries. Let us denote by $P_{\bf r}$ the aforementioned element of the parton gauge group, $M_{\bf+} M_{\bf-}$, but with the $e^{i\vartheta}$ phase factors removed, namely the following unitary operation:

\be\label{Pdeff}
P_{\bf r} f^\dagger_{{{\bf r}}s} P_{\bf r}^{-1}= i\sigma^{y}_{ss'} f_{{\bf r}s'}.
\ee

Thus we see that $P_{\bf r}$ is a gauge particle-hole conjugation that does not change the physical spins. From the existence of this operation, we conclude that for any symmetry which is implemented by acting on the spinons without a particle-hole conjugation, there is another implementation which acts as a particle-hole conjugation, that can obtained by a composition of the original implementation with $P_{\bf r}$. Since such two implementations would differ by the action of a gauge group element, they can be viewed as distinct projective symmetry group (PSG) implementations~\cite{Wen_PSG1,Wen_PSG2} of the same underlying physical symmetry. 

In addition there is a $U(1)$ subgroup of the gauge group acting as:

\be\label{Udeff}
U(\theta_{{\bf r} }) f^\dagger_{{{\bf r} }s} (U(\theta_{{\bf r} }))^{-1}= e^{i \theta_{{\bf r} \tau}} f^\dagger_{{{\bf r} }s}.
 \ee

The gauge transformations in Eqs.~\eqref{Pdeff} and~\eqref{Udeff} are elements of the larger SU(2) parton gauge group of the fermion representation of spins~\cite{Affleck,Dagotto}. We will call {\it pseudoscalar} U(1) spin liquids those states in which there is deconfinement of the U(1) gauge field associated with the above U(1) subgroup and at least one of the underlying symmetries of the problem is implemented on the spinons via a particle-hole conjugation. By following the ideas of PSG~\cite{Wen_PSG1,Wen_PSG2}, it is clear that even within the pseudoscalar $U(1)$ spin liquids, there is a large landscape of possible symmetry enriched phases realizing different PSG implementations of physical symmetries.


\subsection{Symmetries of emergent magnetic fields}


A crucial property distinguishing the pseudoscalar $U(1)$ spin liquids from those with traditional implementation (e.g. Eq.~\eqref{Me}) is that the emergent magnetic field strength, which we will denote by ${\mathcal B}$, transforms in the opposite way to that expected for the ordinary magnetic field experienced by electrons. To show this and to generalize the considerations of the previous section, let us now imagine a general physical symmetry operation, $O$. From our discussion, it follows that there are two kinds of symmetry implementations on the spinons denoted by $O_+$ and $O_-$, and given by:

 \be
\label{Oelect}O_+ f^\dagger_{{{\bf r}}s} (O_+)^{-1}= U^{(+)}_{{\bf r},ss'}({\bf r}) f^\dagger_{{O}{\bf r}s'},\\
O_{-} f^\dagger_{{{\bf r}}s} (O_{-})^{-1}= U^{(-)}_{{\bf r},ss'} f_{{ O}{\bf r}s'},
\ee

 Here ${O}{\bf r}$ is the image of site ${\bf r}$ under $O$ and $U^{(\pm)}_{{\bf r},ss'}$ are unitary matrices. We consider the possibility that the symmetry is anti-unitary, by writing:

 \be
O i O^{-1}= p i, \ p=\pm1,
\ee

where $p=1 (-1)$ when the symmetry is unitary (anti-unitary). Let us now consider a set of trial $U(1)$ spin liquid states that are parametrized by a free fermion Hamiltonian of the form:


\be\label{tdeff}
H[\{ t_{{\bf r},{\bf r'}}^{s,s'}  \}] = \sum_{{\bf r r'}ss'} t_{{\bf r},{\bf r'}}^{s,s'}f^\dagger_{{\bf r} s} f_{{\bf r'} s'}.
 \ee

Here the hopping matrix elements $t_{{\bf r},{\bf r'}}^{s,s'}$ are viewed as a set of variational parameters for the physical trial state, $|\Psi[\{ t_{{\bf r},{\bf r'}}^{s,s'}  \}] \rangle$, which is obtained from the free fermion Slater determinant ground state of Eq.~\eqref{tdeff}, $|\Psi_0[\{ t_{{\bf r},{\bf r'}}^{s,s'}  \}] \rangle$, after the Gutzwiller projection as defined in Eq.\eqref{Gutz}. As discussed in Refs.~\cite{Wen_PSG1,Wen_PSG2}, whenever they are stable against gauge confinement, the above states are expected to describe U(1) spin liquids because the effective spinon Hamiltonian in Eq.~\eqref{tdeff} has a {\it global} U(1) symmetry associated with total spinon number conservation, namely they feature a global U(1) invariant gauge group~\cite{Wen_PSG1,Wen_PSG2}. Importantly, since the Gutzwiller projector by construction always has the same symmetries of the microscopic Hamiltonian written in terms of spin operators, it follows that the Gutzwiller projector commutes with any symmetry implementation $O_{\pm}$. Therefore, in order to guarantee that the physical state is symmetric under the $O$ symmetry, it is sufficient to impose that the free fermion Hamiltonian in Eq.~\eqref{tdeff} is invariant under either one of the symmetry implementations $O_{\pm}$. However, different symmetry implementations will impose different constraints on the variational parameters and thus generally lead to physically distinct states. For unitary symmetries ($p=+$) these constraints are:



\be\label{Mtm}
O_{+} : \ t_{{ O}{\bf r},{O}{\bf r'}}^{s,s'}=\sum_{s_1,s'_1}U^{(+)}_{{\bf r},s_1 s} t_{{\bf r}, {\bf r'}}^{s_1,s'_1} (U^{(+)}_{{\bf r'},s'_1s'})^*,\\
\label{Mte}
O_{-} : \ t_{{ O}{\bf r},{ O}{\bf r'}}^{s,s'}=- \sum_{s_1,s'_1}U^{(-)}_{{\bf r'},s'_1 s'} t_{{\bf r'}, {\bf r}}^{s'_1,s_1} (U^{(-)}_{{\bf r},s_1s})^*.
\ee


While for anti-unitary symmetries ($p=-$) the constraints follow from those above by replacing $t\rightarrow t^*$ in the right hand side. Because the constraints are distinct for the two implementations, $O_{\pm}$, we conclude that the trial physical states will generically be distinct for these two symmetry implementations.

Now, let us imagine that we have found a set of hoppings satisfying one of the above implementations of the $O$ symmetry. Let us refer to the trial state associated with such hoppings as a {\it parent} state. Now, let us consider another mean field state weakly perturbed away from the parent state by adding a small and smooth spatially dependent emergent magnetic field ${\mathcal  B}({\bf x})$. Namely, we consider a new trial state in which the set of hoppings differs from the parent state by adding a trial emergent vector potential ${\mathcal  A}$ that leads to the following change of variational parameters:

\be
t_{{\bf r},{\bf r'}}^{s,s'} \rightarrow e^{i \int_{\bf r'}^{\bf r} d {\bf x} \cdot {\mathcal  A}({\bf x})} t_{{\bf r},{\bf r'}}^{s,s'}.
\ee

Here we are viewing ${\mathcal  A}({\bf x})$ as a function of the continuous coordinate ${\bf x}$ in the ambient space where the lattice model is embedded, and the line integrals are taken along the straight lines joining the initial, ${\bf r'}$, and final, ${\bf r}$, lattice sites associated with the spinon hopping. In order for the new hoppings to satisfy the conditions of Eqs.~\eqref{Mtm}-\eqref{Mte} we can choose the trial vector potentials to satisfy the following conditions:

\be\label{Am}
O_{\pm}:  \ {\bf O} {\mathcal  A}({\bf x})= \pm p {\mathcal  A}({\bf O}{\bf x}),
\ee


where we have assumed ${\bf O}$ is a $2\times 2$ orthogonal matrix. From the above it follows that the emergent magnetic field strength, ${\mathcal  B}=\partial_{\bf x} \times {\mathcal  A}$, would satisfy:

\be\label{Opm}
O_{\pm}: \ {\mathcal  B}({\bf x})=\pm p \det({\bf O}) {\mathcal  B} ( {\bf O}{\bf x}),
\ee

The case of lattice translations (which are unitary) can also be worked out similarly and one obtains two possible implementations $T_{\bf R,\pm}: \ {\mathcal  B}({\bf x})=\pm {\mathcal  B} ({\bf x}+{\bf R})$. In particular, if we consider a uniform trial magnetic field ${\mathcal  B}({\bf x})={\mathcal  B}_0$, we see that emergent magnetic field has the opposite transformation law for the pseudoscalar spin liquids with $O_-$ symmetry implementation, with respect to the traditional U(1) spin liquids with electron-like symmetry implementations, $O_+$. 

In particular, for the mirror, $M$, and $\pi$-rotation, $C$, which are symmetries of $\alpha$-RuCl$_3$ when the physical field is along the {\bf b} axis (see Fig.\Ref{Fig:abc}(b)), it is possible to have pseudoscalar spin liquids with symmetry implementations $M_-$ and $C_-$, which allow for a non-zero average emergent magnetic field ${\mathcal  B}_0$ in their ground state, even though such symmetries forbid the existence of thermal Hall effect, in agreement with experiment~\cite{OngQO}, as we will discuss in more detail in Secs.~\Ref{symmsRuCl}-\Ref{lowEmodel}.


\subsection{Stability of U(1) pseudoscalar spin liquids}
\label{stability}

Let us now discuss whether pseudoscalar U(1) spin liquids are stable phases of matter against gauge confinement or other instabilities that are potentially present in lattice models with emergent gauge fields~\cite{Polyakov,Wenbook,Fradkin}. In the absence of an average magnetic field, pseudoscalar U(1) spin liquids are expected to be subjected to similar kind of stability considerations as ordinary U(1) spin liquids~\cite{Borokhov,Hermele2004,Sung-Sik}. In particular, states with Fermi surfaces are expected to be stable phases of matter~\cite{Sung-Sik}. 



In the presence of an average emergent magnetic ${\mathcal  B}_0$, the spinon Fermi surface would disappear at low energies due Landau quantization. In traditional spinon Fermi surface states where the spinons transform as electrons under symmetry, there will be generically an allowed value of the spinon proper Hall conductivity and consequently an associated allowed Chern-Simons term in the effective low energy action for the emergent gauge fields. This means that ideally at low energies the ordinary spinon Fermi surface states can generically give rise to chiral spin liquids, which can be viewed as the spinon analogues of quantum Hall states and are also stable phases of matter. However, as we argue below, in the case of pseudoscalar U(1) spin liquids a different and curious state of affairs occurs. Namely the symmetries that allow for a non-zero ${\mathcal  B}_0$ can generically forbid the presence of the proper spinon Hall conductivity and consequently also of the Chern-Simons term in the effective low energy action of the gauge fields, and thus these spin liquids do not necessarily evolve into chiral spin liquids at low energies in the presence of ${\mathcal  B}_0$.

To argue for this, we begin by noticing that the Chern-Simons term in the action is proportional to the following product:

\be 
\mathcal{\phi}({\bf x}) \mathcal{B}({\bf x}),
\ee

where $\mathcal{\phi}({\bf x})$ is the scalar part of the emergent gauge field. Since $\mathcal{\phi}({\bf x})$ transforms under symmetries in the same way as the spinon density, $\sum_{s}f^\dagger_{{\bf r} s} f_{{\bf r} s}$, it will be even under an ordinary symmetry implementation, $O_+$, and odd under a pseudoscalar one, $O_-$. Now suppose that the Chern-Simons term was allowed by some ordinary symmetry implementation, $O_+$, which is possible if $\mathcal{B}({\bf x})$ is even under $O_+$. Let us denote the pseudoscalar couterpart of this symmetry by $O_-$. Under $O_-$ both $\mathcal{\phi}({\bf x})$ and $\mathcal{B}({\bf x})$ would now be odd, and therefore the Chern-Simons term would also be allowed. On the other hand, imagine now that certain symmetry with ordinary implementation $O_+$ forbids the Chern-Simons term, which happens when the emergent magnetic field $\mathcal{B}({\bf x})$ is odd under $O_+$. Then in this case under the pseudoscalar counterpart, $O_-$, $\mathcal{B}({\bf x})$ would be even and the scalar potential $\mathcal{\phi}({\bf x})$ would be odd, and therefore the Chern-Simons term will be forbidden as well. In other words, the Chern-Simons term is allowed or forbidden only depending on the physical nature of the underlying symmetry and not on whether it is implemented in an ordinary fashion or a pseudoscalar one on the spinons, in contrast to the emergent field itself $\mathcal{B}({\bf x})$, which might be allowed for one implementation but forbidden for the other. In particular, the Chern-Simons term of 2D U(1) spin liquids satisfies the same symmetry constraints of the physical thermal Hall conductivity $\kappa_{xy}$.

The above leads us to the curious situation that pseudoscalar spin liquids with a symmetry implementation that allows for an average emergent magnetic field $\mathcal{B}_0$, but which forbids the thermal Hall conductivity, $\kappa_{xy}$, and consequently the Chern-Simons term, do not have to generically become chiral spin liquids at low energies. The absence of the Chern-Simons term leads therefore to expect that when the Landau quantization gaps out the spinons while preserving the corresponding symmetry forbidding $\kappa_{xy}$, the leading term in the effective gauge action after integrating out the spinon fields will be the Maxwell term. However, due to the ubiquitous confinement expected for ordinary 2D compact QED~\cite{Polyakov,Wenbook,Fradkin}, the true low energy ground state would be a confined phase, and then generically the ultimate low energy state will display some form ordinary spontaneous symmetry breaking (or some symmetry preserving paramagnetic phase if the symmetries allow it). If one could artificially make the spinon coupling to gauge fields weak, this instabilities would occur at low energy scales typically below the effective cyclotron energy of the parent pseudoscalar spinons Fermi surface state, and might lead to a rather rich and delicate sequence of phase transitions (spinon-gauge coupling is however generically strong). To picture this more intuitively the reader could, very roughly speaking, imagine the formation of stripe or bubble phases in high Landau levels~\cite{LLstripebubbles} as analogues of these instabilities of the pseudos-scalar spin liquids with forbidden Chern-Simons terms, except that rather than having long-wavelength charged ordered structures the spinons would have some order of the spin densities. Curiously, a somewhat reminiscent phenomenon is also encountered in the case of strong unidirectional spin density wave states subjected to magnetic fields, which display some form of quantum oscillations that are expected to evolve into a sequence of phase transitions at low temperatures~\cite{Gorkov,Chaikin}. 

\section{Pseudoscalar U(1) spin liquids in $\alpha$-RuCl$_3$}
\label{pseudoRuCl}

We will now apply the ideas described in Sec.~\ref{Pseudo-general} to $\alpha$-RuCl$_3$. We begin by reviewing some of the general properties of this material. Each layer of $\alpha$-RuCl$_3$ contains a honeycomb lattice of RuCl octahedra (see Fig.~\Ref{Fig:abc}(a)). A Hilbert space with one effective spin-1/2 degree of freedom per octahedron and a Hamiltonian with exchange couplings ranging up to about $\sim 10meV$~\cite{Khaliullin2009,Valenti_2016,Valenti_2017,Takagi_2019,Chernyshev}, are believed to describe well its low energy properties. Crucially, the material is a good electrical insulator with transport and optical gaps ranging on the order of $\sim0.1eV - 2eV$~\cite{Kim_2014,Burch_2016,Park_2016}. We will therefore imagine that the physics emerges from a spin  Heisenberg-like model of the form:

\be\label{Hgen}
H_{J}+H_{Z}=\sum_{all} J^{\alpha\alpha'}_{{\bf r},{\bf r'}} \sigma^\alpha_{{\bf r}} \sigma^{\alpha'}_{{\bf r'}}-\mu^{\alpha \alpha'} B^\alpha \sigma^{\alpha'}_{{\bf r}}.
\ee

Here $\sigma^\alpha_{{\bf r}}$ denotes the $\alpha  \in \{x,y,z\}$ Pauli matrix at honeycomb site ${\bf r}$. $J^{\alpha\alpha'}_{{\bf r},{\bf r'}}$ and $\mu^{\alpha \alpha'}$ are exchange couplings and magnetic moment tensors, and $B^\alpha$ is the experimentally applied magnetic field along axis $\alpha$. We will take the spin axes to be aligned with the crystal axes as follows: $\{x,y,z\} \leftrightarrow \{{\bf a,b,c}'\}$ (see Fig.~\Ref{Fig:abc}).

\subsection{Symmetries of $\alpha$-RuCl$_3$}
\label{symmsRuCl}

The honeycomb lattice of $\alpha$-RuCl$_3$ is slightly contracted along the ${\bf b}$ axis (see Fig.~\Ref{Fig:abc}(a)). In order to understand the implications of symmetries it is useful to imagine an ideal model with all the symmetries of the un-contracted lattice and view the terms that break these as smaller corrections. Namely, we partition the exchange part of the Hamiltonian into:

\be
H_{J}=H_{J}^{(0)}+H_{J}^{(1)}.
\ee

The symmetry group of $H_{J}^{(0)}$ contains the following operations:

\be\label{symms}
\{M_{\bf b}, M_{\bf c'} C_{6{\bf  c'}}, \mathcal {T},T_{{\bf R}}\} \rightarrow \{M, R , \mathcal {T},T_{{\bf R}}\}.
\ee

Here $M_{\bf h}$ denotes a mirror that leaves the ${\bf h}$ axis invariant, $C_{n {\bf h}}$ is the rotation by an angle $2\pi/n$ around axis ${\bf h}$, $\mathcal {T}$ is the usual anti-unitary time reversal ($\mathcal {T}^2=-1$), $T_{{\bf R}}$ is a short-hand for the honeycomb lattice translations $T_{n_1{\bf R}_1+n_2{\bf R}_2}$ associated with the Bravais lattice vectors ${\bf R}_{1,2}$, and the symbols following the ``$\rightarrow$" are simplified notations for future convenience. All of the above symmetries are viewed as acting in the three-dimensional space embedding the two-dimensional honeycomb lattice. The more realistic Hamiltonian $H_{J}$ for the contracted lattice has all of the above except the roto-reflection $R=M_{\bf c'} C_{6{\bf  c'}}$. There is however direct experimental evidence that the breaking of this roto-reflection is weak, from detailed angular dependence measurements of the specific heat~\cite{Shibauchi}.  Now, in the presence of an externally applied field ${\bf B}$ along the ${\bf b}$ axis, the full Hamiltonian, $H_{J}+H_{Z}$, has  the following remnant symmetries:

\be\label{B||b}
{\bf B} \parallel {\bf b}: \ \{M, C, T_{{\bf R}}\}.
\ee

where $C$ denotes the $\pi$-rotation $C_{2{\bf  b}}$ which can be expressed in terms of the original symmetries as follows:

\be\label{CMR}
C =M R^3.
\ee

The thermal Hall conductivity is odd with respect to either $M $ or $C $ operations and would therefore vanish if at least one of these symmetries remains unbroken in the ground state. In contrast when the external field is along the ${\bf a}$ axis, the remnant symmetries are:

\be\label{B||a}
{\bf B} \parallel {\bf a}: \ \{M \mathcal {T}, C \mathcal {T}, T_{{\bf R}}\}, 
\ee

However the thermal Hall conductivity is even under all of the above operations and will therefore generally be present in this case. Finally, the zig-zag AFM state in the presence of fields along the ${\bf b}$ axis~\cite{Coldea_2015,Nagler_2016,Nagler_2018,Nagler_2019}, is expected to break spontaneously the symmetry group from Eq.~\eqref{B||b} down to: 


\be\label{AFMb}
\{M , C T_{{\bf R}_1}, T_{n_1{\bf X}_1+n_2{\bf X}_2}\} \rightarrow \{M , C T_{{\bf R}_1}, T_{{\bf X}}\}.
\ee

where ${\bf X}_{1}={\bf R}_1-{\bf R}_2$, ${\bf X}_{2}={\bf R}_1+{\bf R}_2$ are the new Bravais lattice vectors translating the enlarged magnetic unit cells (see Fig.~\Ref{Fig:abc}(d)) and the symbols after ``$\rightarrow$" are simplified notations. The thermal Hall conductivity is odd under either the $M$ or the nonsymorphic $C T_{{\bf R}_1}$, explaining its absence in experiments~\cite{OngQO,Matsuda_2020}. And when the field is along the ${\bf a}$ axis the zig-zag AFM is expected to contain the following operations:


\be\label{AFMa}
\{M\mathcal {T}T_{{\bf R}_1} , C\mathcal {T} , T_{{\bf X}}\}.
\ee


\subsection{pseudoscalar spinon Fermi surface state with the symmetries of $\alpha$- RuCl$_3$}
\label{examplepseudo}

In this section we will present a specific example of a pseudoscalar spinon Fermi surface state with all the symmetries of $\alpha$-RuCl$_3$. We caution that this example is simply meant to illustrate the previous general ideas in a concrete setting. There is a wide land-scape of symmetry enriched states even within each of the scenarios we have previously outlined and also uncertainties in the precise microscopic model of $\alpha$-RuCl$_3$, hindering a detailed energetic analysis

Moreover, the very nature of the state that is realized in experiments suggests that its precise energetics is hard to capture microscopically. This is in part because the emergent magnetic field presumably plays an important role in tilting the energetic balance in favor of the pseudoscalar spin liquid states. However, because its strength can adjust as a continuous variable the system can have very large and variable magnetic unit cells comprising many atomic unit cells. There is also no expected simple commensurability between the size of the spinon Fermi surface and the Brillouin zone. In fact, if one performs a naive estimate of the area of the Fermi surfaces in experiment~\cite{OngQO}, by naively assuming that magnitude of the emergent magnetic field equals the in-plane physical field, one obtains a Fermi surface area of about $\sim 0.3 \% $ the size of the Brillouin zone. This is a very rough estimate, but it is a compelling indication that the Fermi pockets are small and the spinons are dilute compared to lattice spacing (the typical inter-spinon distance would be $\sim 10nm$). Another way to state this is as follows: suppose we would like that the spinons have a Fermi surface that is half of the $\alpha$- RuCl$_3$ Brillouin zone. Then, in order to match the experimental QO's period~\cite{OngQO}, one would need to assume that the effective constant, $\alpha$, controlling the proportionality between the magnitude of the emergent field, ${\mathcal B}_0$, and the physical in-plane field ${\bf B}$:

\be
|{\mathcal B}_0| \approx \alpha |{\bf B}|, 
\ee

is about $\alpha \approx 10^3$. This would be a tremendous enhancement of the amount of orbital flux experienced by spinons, notwithstanding the fact that the applied field is strictly in-plane. Therefore, it seems more natural to assume that the pockets are relatively small compared to the Brillouin zone while the enhancement of flux is not so gigantic. All of the above indicates that a good portion of the energetics behind the precise Fermi surface ground state realized experimentally is ``long-wavelength", and therefore hard to capture in numerical studies of small system sizes or in un-controlled numerical or analytical calculations. Moreover, as we discussed in Sec.~\ref{stability}, the pseudoscalar spinon Fermi surface states have a stronger tendency to form broken symmetry states in their ultimate low energy ground states in the presence of an average magnetic field, due to the absence of the spinon Chern-Simons term which generically leads to non-chiral spin liquids.


\begin{table}
\centering
\renewcommand{\arraystretch}{1.3}
\begin{tabular}{|c| c| c| c|}
\hline 
Symmetry &Symmetry action& ${\mathcal  B}_0$ & $\kappa_{xy}$ \\[0.5ex] 
\hline 
$M_-$ & $M_- f^\dagger_{{\bf r} s} M_-^{-1}= \tau^z_{\bf r} f_{M{\bf  r} s} $ & + & -  \\ 
\hline 
$R_+$ & $R_+ f^\dagger_{{\bf r} s} R_+^{-1}=- i \sigma^z_{s}  z^{\sigma^z_{s}/4} f^\dagger_{R{\bf  r} s} $ & + & + \\    \hline 
${\mathcal T}_+$ & ${\mathcal T}_+ f^\dagger_{{\bf r} s} {\mathcal T}_+^{-1}= i \sigma^y_{ss'}  f^\dagger_{{\bf r} s'} $ & - & -  \\ \hline
$T_{{\bf R}+}$ & $T_{{\bf R}+} f^\dagger_{{\bf r} s} T_{{\bf R}+}^{-1}= f^\dagger_{{\bf r+R} s} $ & + & +  \\ \hline 
$C_-=M_- (R_+)^3$ & $C_- f^\dagger_{{\bf r} s} C_-^{-1}= \tau^z_{\bf r} f_{C{\bf  r} s} $ & + & -  \\ \hline 
\end{tabular}
\caption{Symmetry action on the spinon operator $f^\dagger_{{\bf r} s}$ at honeycomb site ${\bf r}$ and spin $s$. $\tau^z_{\bf r}=1 (-1)$ when ${\bf r}$ belongs to the A (B) sublattice and $z=e^{i 2 \pi/3}$. The last columns denote by + (-) whether the emergent field, ${\mathcal  B}_0$, and the thermal Hall conductivity, $\kappa_{xy}$, are allowed (forbidden). The sub-script $+/-$ in the symmetry operation denotes whether it is implemented in an ordinary electron-like fashion ($+$) or in a ``pseudoscalar" fashion ($-$) involving a spinon particle-hole conjugation. Only $\{M_-, C_-, T_{{\bf R}+}\}$ remain symmetries with a field along the ${\bf b}$ axis.
\label{tsymm}}
\end{table}

With the above caveats in mind, we would like, however, to present a concrete toy illustration of a pseudoscalar spinon Fermi surface state in this section, consistent with all the symmetries of $\alpha$-RuCl$_3$. To do this we begin by considering the following possible symmetry implementations for the ideal fully symmetric Hamiltonian, $H_{J}^{(0)}$, with symmetries listed in Eq.~\eqref{symms}:

\be\label{parent3B}
 \ \{M_- , R_+, \mathcal{T}_+,T_{{\bf R}+}\}.
\ee

Where sub-script ``-" indicates that the symmetry is pseudoscalar, namely that it acts as particle-hole conjugation, and ``+" indicates that is does not act as a particle-hole conjugation on the spinons. From Eq.~\eqref{Opm} it follows that the emergent magnetic field, ${\mathcal  B}_0$, is only odd and hence forbidden by $\mathcal{T}_+$, but allowed by all of the other symmetries. Here we will concentrate on the situation when the physical field is along axis ${\bf b}$, which tends to impose more constraints on the physics, although related considerations can be developed when the field is along axis ${\bf a}$. Thus in the presence of the Zeeman field along ${\bf b}$, the symmetries are lowered to those listed in Eq.~\eqref{CMR} and, therefore, we have the following remnant symmetries acting on the spinons:

\be\label{B||b2}
{\bf B} \parallel {\bf b}: \ \{M_-, M_- R_+^3, T_{{\bf R}+}\}\rightarrow \{M_-, C_-, T_{{\bf R}+}\}.
\ee

Where the symbols after the $``\rightarrow"$ are short-hand notations. From the experimental absence of $\kappa_{xy}$ one concludes that at least one of the two symmetries $M$ or $C$ is present but not necessarily both. It is in principle possible that one of them is spontaneously broken in the spin liquid state, but we will assume that the spin liquid respects all the symmetry of the Hamiltonian. Thus we see that both these symmetries force $\kappa_{xy}=0$, but their pseudos-scalar implementations allow for a finite ${\mathcal  B}_0$ and thus the existence of quantum oscillations, in agreement with experiment~\cite{OngQO}. Moreover, notice that the time reversal operation $ \mathcal{T}_+$, which is the only symmetry which forbids ${\mathcal  B}_0$, is broken {\it explicictly} by the external in-plane field. This can be viewed as part of the mechanism allowing to understand the growth of the emergent magnetic field ${\mathcal  B}_0$ with increasing applied in-plane fields seen in experiment~\cite{OngQO}.

To be able to define a simple notion of spinon dispersion, we will consider a parent state with zero flux per honeycomb unit cell and ${\mathcal  B}_0=0$. The state realized in experiments is then viewed as a perturbed version of this parent state with  a finite but small ${\mathcal  B}_0$. We choose the specific action for the symmetries to be that listed in Table~\ref{tsymm}. This symmetry action dictates via Eqs.~\eqref{Mtm}-\eqref{Mte} the allowed hopping terms, $\{ t_{{\bf r},{\bf r'}}^{s,s'}  \}$, parametrizing the variational wave-function of Eq.~\eqref{Gutz}. It is convenient to organize these hoppings by their spatial range in the lattice, since shorter range hoppings are typically dominant. The leading terms are the on-site terms. If one imposes all the symmetries from Table~\ref{tsymm} there is no allowed on-site term. However for the lower symmetry group that is present when the external field is along the $\bf{b}$ axis (see Eq.~\eqref{B||b2}) the following on-site term becomes allowed:

\be\label{onsitet0}
H[\{ t_{{\bf r},{\bf r'}}^{s,s'}  \}] = - \sum_{{\bf r}ss'} t_{0\perp} \sigma^y_{s,s'}f^\dagger_{{\bf r} s} f_{{\bf r} s'}.
\ee

\begin{figure}[!t]
\begin{center}
\includegraphics[width=0.45\textwidth]{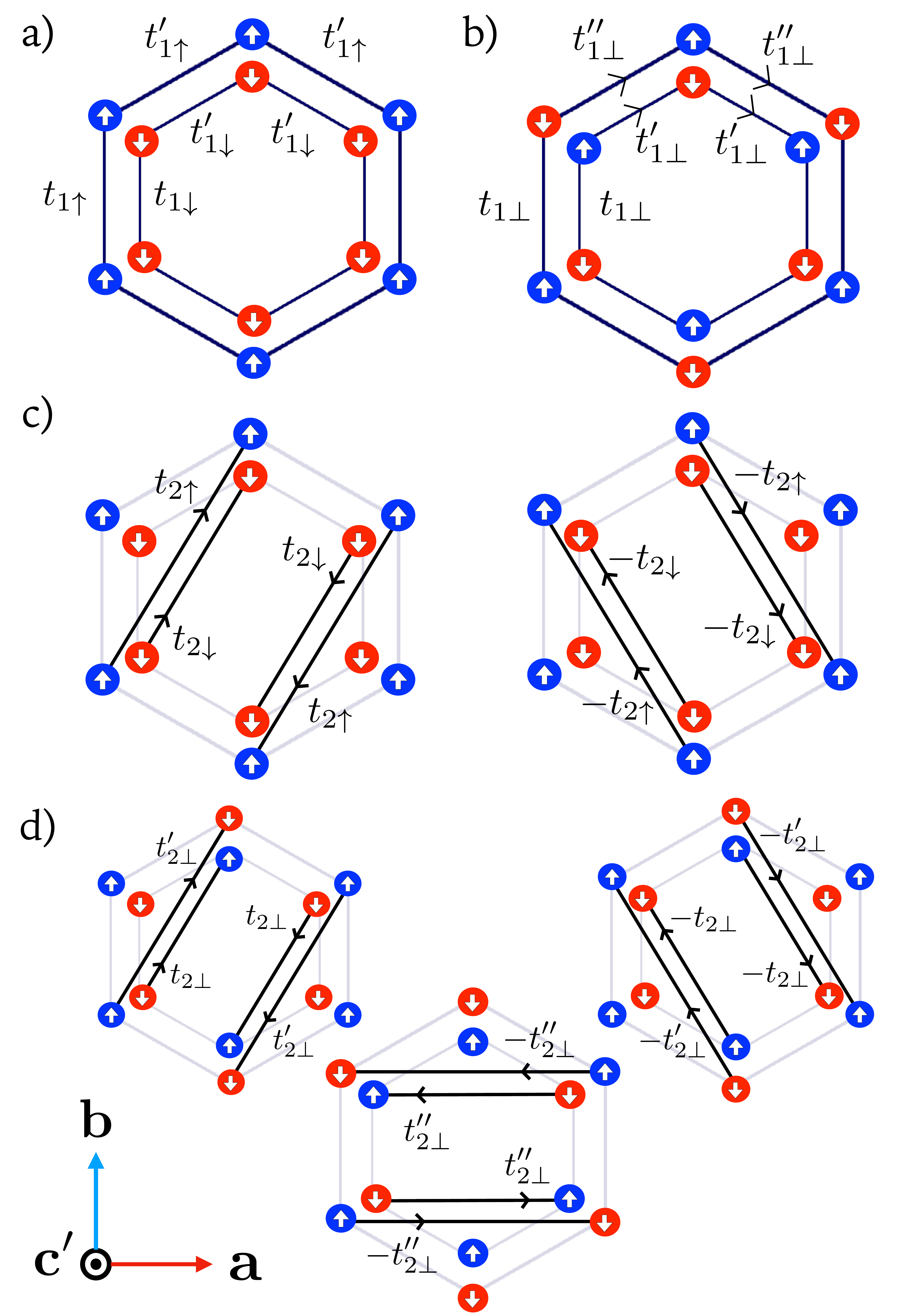}
\end{center}
\par
\caption{(Color online) Symmetry allowed spinon hoppings for an in-plane field along the ${\bf b}$ axis. Solid lines without arrows denote real hoppings, while directed lines denote complex hoppings. Nearest neighbor spin-preserving (a)  and spin-flipping (b) hoppings. Second neighbor spin-preserving (c) and spin-flipping (d) hoppings.}
\label{Fig:ts}
\end{figure}

Even though the above term has the same form as a Zeeman field along the $\bf{b}$ axis, the parameter $t_{0\perp}$ should not be viewed in general as the bare Zeeman coupling entering the microscopic Hamiltonian from Eq.\eqref{Hgen}, but as a variational parameter to be optimized to minimize the total energy. Using Eqs.~\eqref{Mtm}-\eqref{Mte} one can in this way continue finding the nearest and further neighbor spin dependent hoppings, to enlarge the allowed parameter space describing the variational wavefunction. The symmetry allowed hoppings up to second neighbors are shown in Fig.~\Ref{Fig:ts}. There are four independent and real nearest-neighbor spin-preserving hoppings $\{t_{1\uparrow},t'_{1\uparrow},t_{1\downarrow},t'_{1\downarrow}\}$ allowed by symmetries from Eq.~\eqref{B||b2} when the field is in-plane (see Fig.~\ref{Fig:ts}(a)). In the absence of external in-plane field, namely under the full symmetry group from Eq.\eqref{parent3B}, these four terms collapse onto a single independent real parameter:

\be\label{t1up}
t_{1\uparrow}=t'_{1\uparrow}=t_{1\downarrow}=t'_{1\downarrow}.
\ee

There is one real $t_{1\perp}$ and two complex $\{t'_{1\perp},t''_{1\perp}\}$ spin-flipping nearest neighbor hoppings when the field is along the ${\bf b}$ axis (see Fig.~\ref{Fig:ts}(b)). Upon the imposing the full symmetry group from Eq.\eqref{parent3B} these five independent real parameters collapse onto a single one $t_{1\perp}$, as follows:

\be\label{t1perp}
t'_{1\perp}=z t_{1\perp}, \ t''_{1\perp}=z^{-1} t_{1\perp}, \ z=e^{i 2 \pi/3}.
\ee

There are two complex second-neighbor spin-preserving hoppings $\{t_{2\uparrow},t_{2\downarrow}\}$ when the field is along the ${\bf b}$ axis (see Fig.~\ref{Fig:ts}(c)). These hoppings are actually forbidden when the full symmetry group from Eq.\eqref{parent3B} is imposed, namely $t_{2\uparrow}=t_{2\downarrow}=0$. Lastly, there are three complex second-neighbor spin-flipping hoppings $\{t_{2\perp},t'_{2\perp},t''_{2\perp}\}$ when the field is along the ${\bf b}$ axis (see Fig.~\ref{Fig:ts}(c)). Upon the imposing the full symmetry group from Eq.\eqref{parent3B} these six independent real parameters collapse onto a single one $t_{2} \in {\mathbb R}$, as follows:

\be
t_{2\perp}=z^{-1/2} t_{2},  t'_{2\perp}=z^{-1} t_{2}, t''_{2\perp}=t_{2},
\ee

\begin{figure}[!t]
\begin{center}
\includegraphics[width=0.47\textwidth]{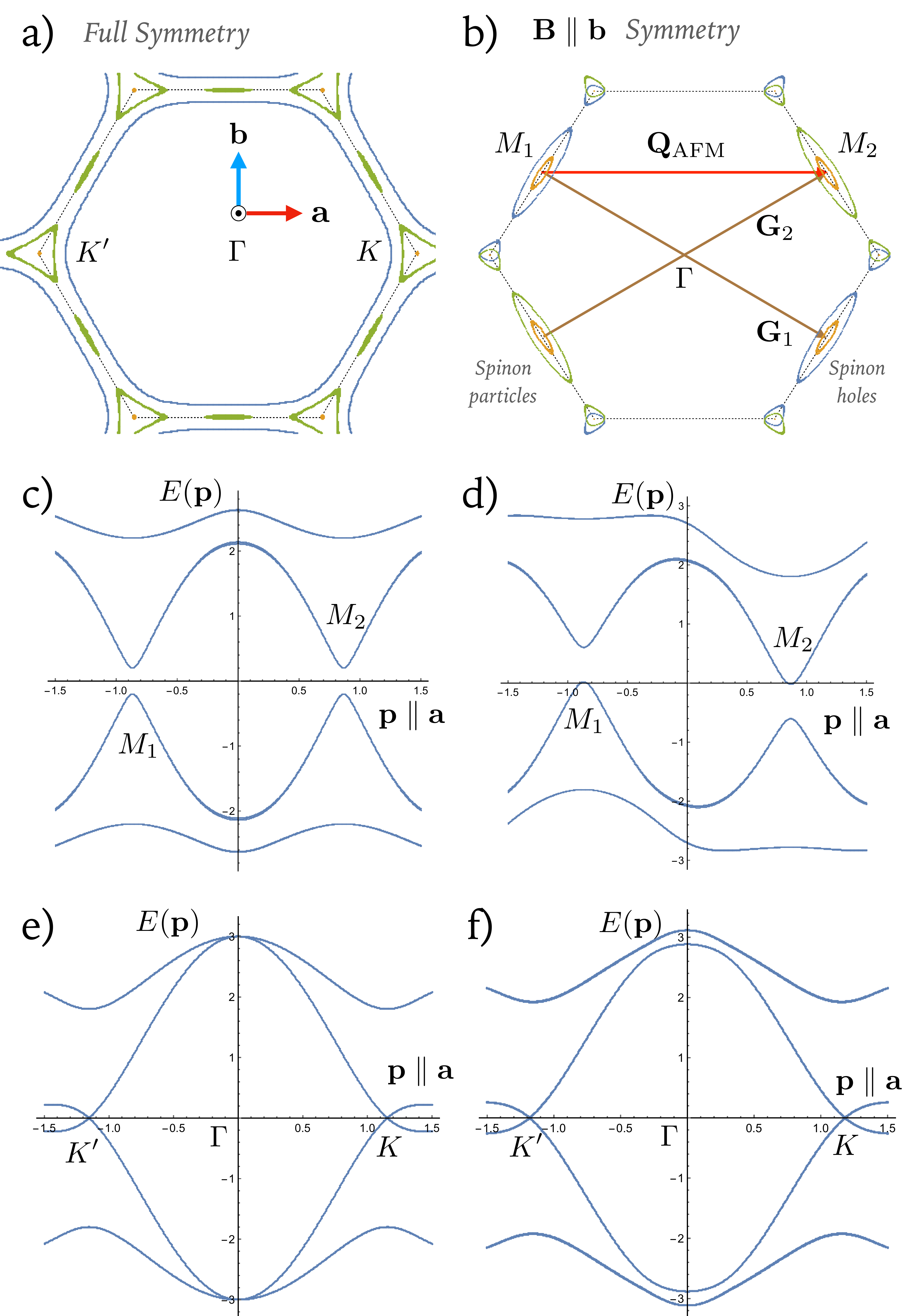}
\end{center}
\par
\caption{(Color online) (a) Spinon dispersion iso-energy contours at $E=0$ (orange Dirac points at $K$ and $K'$), $E=0.2 t_{1\uparrow}$ (green) and $E=0.4 t_{1\uparrow} $ (blue), and dispersions (c) and (e) for the state without in-plane field with all the symmetries of Table~\Ref{tsymm}. The non-zero spinon hoppings are $t_{1\uparrow}$ (unit of energy) (from Eq.\eqref{t1up}), and $t_{1\perp}=0.6 t_{1\uparrow}$ (Eq.\eqref{t1perp}). (b) Spinon Fermi surfaces for the lower symmetry state (in-plane field along the ${\bf b}$ axis) located at $M_1$ and $M_2$ points (orange contours) coexisting with Dirac nodes (orange dots at $K$ and $K'$) and iso-energy contours at $E=0.1 t_{1\uparrow}$ (green) and $E=-0.1 t_{1\uparrow}$ (blue). Cuts of the dispersions are shown in (d) and (f). The non-zero spinon hoppings are $t_{1\uparrow}=t'_{1\uparrow}=t_{1\downarrow}=t'_{1\downarrow}$ (from Eq.\eqref{t1up}),  $t_{0\perp}=0.1 t_{1\uparrow}$ (Eq.\eqref{onsitet0}), $t_{1\perp}=0.6  t_{1\uparrow}$, $t'_{1\perp}=0.66  t_{1\uparrow} z$, $t'_{1\perp}=0.66  t_{1\uparrow} z^{-1}$ (Eq.\eqref{t1perp}), $t_{2\uparrow}=t_{2\downarrow}=0.1 t_{1\uparrow}$ (see Fig.~\ref{Fig:ts}). We have verified that properties are robust against small changes around the above parameters.}
\label{Fig:dispersions}
\end{figure}

Here we use the convention $z^{-1/2}=e^{-i \pi/3}$. Figure~\ref{Fig:dispersions} shows the dispersion of the spinons for some specific choice of parameters for the full symmetry from Eq.\eqref{parent3B} and for some other choice with the lower symmetry corresponding to ${\bf B} \parallel {\bf b}$ from Eq.~\eqref{B||b2}. The parameters for this figure have been adjusted by hand so as to have small Fermi surfaces located at the $M$ points, because these locations are consistent with the instability of the spinon Fermi surface into the zig-zag AFM, as we discuss next. We also note that for this choice of parameters the Fermi surfaces at M points coexist with Dirac nodes at the K points (see Fig.~\ref{Fig:dispersions}(a),(f)).

The zig-zag AFM is believed to break the translational group of the honeycomb producing an enlarged unit cell with four inequivalent sites and a new rectangular Bravais lattice generated by lattice vectors ${\bf X}_{1,2}$ (see Fig.\ref{Fig:abc}(d)):

\be
{\bf X}_{1}={\bf R}_1-{\bf R}_2, {\bf X}_{2}={\bf R}_1+{\bf R}_2.
\ee

This can be achieved by starting from the pseudoscalar U(1) parent and bose-condensing a spinon particle-hole bilinear which reduces the symmetry down to that of the zig-zag AFM listed in Eq.~\eqref{AFMb}. Such condensing boson operators can be viewed as a new term added to the bilinear spinon Hamiltonian from Eq.~\eqref{tdeff} that parametrizes the trial state. The scattering vector associated with such reduction of translational symmetry is shown in Fig.~\ref{Fig:dispersions}(b) and is given by:

\be\label{QAFM}
{\bf Q}_{\rm AFM}=\frac{{\bf G}_1+{\bf G}_2}{2},
\ee

where ${\bf G}_{1,2}$ are the reciprocal Bravais lattice vectors of the honeycomb (see Fig.~\Ref{Fig:dispersions}(b)). The allowed leading terms describing the zig-zag AFM are simple onsite terms such as that from Eq.\eqref{onsitet0} except that they break the translational symmetry and have the form of the local staggered Zeeman fields associated with the spin moments in the zig-zag AFM, exactly like those depicted in Fig.~\ref{Fig:abc}(d). If the spinon Fermi surface is fully gapped as a result of such condensation, the resulting phase will generically have strong confinement for the U(1) gauge field and the spinons will be strongly bound with a linear potential~\cite{Polyakov,Wenbook,Fradkin}. If the spinon Fermi surfaces do not fully gap as a result of the condensation of the spinon particle-hole bilinear, one would have a state with coexistence of AFM order and reconstructed spinon Fermi surfaces. As a result of the partial gapping of the Fermi surface, this state would naturally have Fermi surfaces enclosing a smaller area, and therefore would naturally have smaller frequencies of oscillation (see subsequent discussion in Sec.~\ref{lowEmodel}). Therefore this partially gapped spinon Fermi surface coexisting with the AFM order is a plausible candidate for the intermediate AFM state~\cite{Vojta,Nagler2021}, which gives rise to different oscillations at lower fields with smaller frequencies, as summarized in {\bf Observation 2} from Sec.\ref{intro}. 

One important caveat is that the rationalization of the emergence of the zig-zag AFM state via a particle-hole instability from the spin-liquid parent state relies on weak coupling intuition. Namely it visualizes a hypothetical dispersion of non-interacting spinons and imagines adding spinon interactions as a perturbation. However, in the current context the spinons are inherently strongly coupled and there is no clear way to deform the problem to turn off their interactions. Nevertheless, if the transition in a realistic model of $\alpha$-RuCl$_3$ is proximate to an ideal continuous transition from a parent spin liquid state to the zig-zag AFM, then it is reasonable to expect that near such transition there are spinon pockets that are connected via the ${\bf Q}_{\rm AFM}$ wave-vectors. This is, however, not necessarily the only possibility. One could for example imagine an alternative scenario in which the spinon pockets are at other locations of the Brillouin zone that are not connected by the ${\bf Q}_{\rm AFM}$ vector, and the transition into the zig-zag AFM proceeds via a condensation of another bosonic spinon particle-hole mode that has wave-vector ${\bf Q}_{\rm AFM}$ but that is not naturally associated with the gapless excitations from the spinon Fermi surfaces. 

Notice, in connection to the above discussion, that the spinon dispersion illustrated in Fig.~\ref{Fig:dispersions}(a),(c),(e) with all the symmetries of Table~\ref{tsymm} in the absence of in-plane field, does not display any spinon Fermi surfaces or gapless spinon nodal points with the nesting that would give rise to the zig-zag AFM. This is not necessarily an issue since in the absence of in-plane field the parent spin liquid is not supposed to be a very close energetic competitor to the true ground state zig-zag AFM. As described above, if indeed the transition proceeds continuously via the spinon Fermi surface particle-hole instability picture, then the spinon pockets only need to be connected by ${\bf Q}_{\rm AFM}$ for the parameters that describe the state proximate to the transition. These dispersions in Fig.~\ref{Fig:dispersions}(a),(c),(e) feature, however, low lying spinon particle-hole excitations with a small gap and wave-vector ${\bf Q}_{\rm AFM}$, which can be viewed as remnants of the energetic proximity of the parent spin liquid state deeper into the zig-zag anti-ferromagnetic state.


\section{Simplified low energy models and Quantum Oscillations of pseudoscalar spinon Fermi surfaces}
\label{lowEmodel}

As we have seen the pseudoscalar spinon Fermi surface states feature pairs of particle and hole spinon pockets that are related by the space symmetries which act as particle-hole conjugations. Let us consider a simple scenario in which each of these individual pockets is small and can thus be approximated as an ellipse. When the field is along the ${\bf b}$ axis  we have two particle-hole conjugating symmetries $M_-$ and $C_-$ described in Table~\ref{tsymm}. The action of these symmetries in crystal momentum basis is:

\be
M_- f^\dagger_{{\bf p},\tau_z s} M_-^{-1}= \tau_z f_{-M{\bf p},\tau_z s}.
\ee

Where ${\bf p}$ is the crystal momentum in the first Brillouin zone of the honeycomb lattice, $\tau_z=\pm 1$ denotes the A and B sublattices, and the action for $C_-$ can be obtained by replacing $M\rightarrow C$ above. Notice that because of their particle-hole conjugating nature these symmetries relate states at ``minus" the momenta expected for ordinary symmetry action. Generically when a Fermi surface is not centered at a special momenta in the Brillouin zone one expects four pockets related by these symmetries. Let us denote the crystal momenta location of these pockets as ${\bf P}_i$, with $i=\{1,...,4\}$. Take the pocket $i=1$ to be particle-like, its low energy dispersion will be:

\be
E_1({\bf p+P}_1)\approx \frac{({\bf p\cdot n}_1)^2}{2 m_1}+ \frac{({\bf p\cdot n}_{1\perp})^2}{2 m_2}-E_0.
\ee 

Where ${\bf n_1}$ and ${\bf n}_{1\perp}\equiv \hat{\bf z} \times {\bf n_1}$ are the two principal axes of the ellipse with respective masses $\{m_1,m_2\}$ and $E_0$ controls the size of the Fermi surface. The momenta and paramaters of the other pockets follow from the $C_-$ and $M_-$ action. Specifically the other two hole-like pockets and their principal axes are:

\be
{\bf P}_2 =- C {\bf P}_1, \ {\bf P}_3 =- M {\bf P}_1=-{\bf P}_2, \\
{\bf n}_2 =- C {\bf n}_1, \ {\bf n}_3 =- M {\bf n}_1=-{\bf n}_2.
\ee 

The pocket 2 has hole-like dispersion given by:

\be
E_{2}({\bf p+P}_{2})\approx -\frac{({\bf p\cdot n}_2)^2}{2 m_1}-\frac{({\bf p\cdot n}_{2\perp})^2}{2 m_2}+E_0,
\ee

And similarly pocket 3 has an identical dispersion to pocket 2 but centered at ${\bf P}_{3}$. On the other hand the fourth pocket is particle-like and is located at:

\be
{\bf P}_4 = C M {\bf P}_1= -{\bf P}_1.
\ee

And the principal axes and the dispersion of this pocket are the same as in pocket 1 but centered at ${\bf P}_4$. 

On the other hand, when the momenta of the particle-like pockets 1 and 4 differ by a reciprocal Bravais lattice wave-vector they are then just one and the same pocket which is left invariant by the inversion symmetry implemented by the product $C_-M_-$, and similarly the hole-like pockets 2 and 3 are also only a single pocket. There are only two special locations in the Brillouin zone of the honeycomb lattice that satisfy this, namely the two M points depicted in Fig.~\ref{Fig:dispersions}(b). More precisely, these are the only crystal momenta that equal minus themselves up to a reciprocal vector, and are thus invariant under the product $MC$, but which are not invariant under the individual $C$ and $M$ symmetries. Notably these special M points are exactly connected by the wave-vector associated with the spinon particle-hole instability that would drive the formation of the zig-zag AFM state and are the locations of the Fermi pockets for the example discussed in~\ref{examplepseudo} and depicted in Fig.~\ref{Fig:dispersions}. Therefore this scenario is nicely consistent with the phenomenology of $\alpha$-RuCl$_3$.

Let us describe now in more detail the quantum oscillations for both of the scenarios of either four or two small elliptical pockets described above. The Landau levels of spinons that form in the presence of the emergent field ${\mathcal B}_0$, come in pairs of positive and negative energies and are given by:

\be
E_{i,n} = s_i (\omega_c (n+1/2)-E_0),
\ee

where $i$ labels the elliptical pockets, $s_i=+1(-1)$ for particle (hole) pockets, $\omega_c=|{\mathcal B}_0| / \sqrt{m_1 m_2}$ is the spinon cyclotron energy scale, and $n=0,1,2...$ is the cyclotron index of each landau level within a pocket. One useful difference between the pseudoscalar spinons and ordinary electrons or scalar spinons is that the chemical potential is fixed to be zero in the former case due to the particle-hole symmetries. In the case of electrons or scalar spinons the chemical potential can oscillate in the physical system which typically are better approximated by an ensemble with fixed density, and this tends to complicate the precise analysis of quantum oscillations~\cite{Harrison,Itskovsky,Champel,Champel2}. We will also assume here for simplicity that the system has a fixed ${\mathcal B}_0$, although we also expect the same kind of behavior described in Refs.~\cite{Motrunich,SodemannPRB}, in which ${\mathcal B}_0$ adjusts itself energetically and it is not directly experimentally fixed. Such behavior can lead to complex stair-case dependence of ${\mathcal B}_0$ as a function of the experimentally controlled physical field, and also to the possibility of multiple metastable states with slightly different values of ${\mathcal B}_0$ as described in Refs.~\cite{Motrunich,SodemannPRB} at low temperatures. This behavior could be in part related to the complex hysteric behavior of the thermal conductivity reported in Ref.~\cite{OngQO} at the lowest temperatures. Such behavior tends to disappear when the temperature exceeds the cyclotron energy~\cite{SodemannPRB}, and therefore our current treatment can be justified in this regime.
 
We follow Ref.~\cite{streamlined2doscillations} which summarizes Shoenberg's derivation of two-dimensional quantum oscillations~\cite{Shoenberg}. We begin by writing the spinon DOS per unit energy per unit area as:

\be
\rho  (\epsilon) =   \frac{|{\mathcal B}_0|}{2\pi} \sum _{i} \sum _{n=0}^{\infty } D_0 (\epsilon-E_{i,n}),
\ee

Here $D_0(\epsilon)$ is a Lorentzian function normalized to unity and width $\Gamma$ that captures the broadening of Landau levels arising e.g. from disorder effects. Now, in the usual limit in which $\omega_c \ll E_F=E_0$, the lower bound of the sum over $n$ can be replaced as $0 \rightarrow -\infty$, so as to approximate the oscillatory density of states as a strictly periodic function of energy. The constant part of this function can also be dropped as it contributes only to the non-oscillatory background. For fermions with fixed chemical potential $\mu=0$, the free energy is:

\be
\Omega=-\beta^{-1} \int d \epsilon \rho  (\epsilon) \log \left(1+e^{-\beta \epsilon}\right),
\ee

From which one obtains the following form for the oscillatory part of this free energy as function of $1/|{\mathcal B}_0|$~\cite{streamlined2doscillations}:

\be\label{Omegaosc}
\Omega_{\rm osc} = \frac {g\omega_c | {\mathcal B} _ 0 |} {2 \pi^3}\sum_ {k = 1}^{\infty} L_k D_k \frac{(-1)^{k+1}}{k^2} \cos \left(\frac{kS}{|{\mathcal B}_0|}\right).
\ee

where $S=\pi p_{F1} p_{F2}=2 \pi \sqrt{m_1 m_2} E_0$ is the Fermi surface area of a single spinon pocket (all have them have the same area in our elliptical models), $g$ is the total number of pockets (namely $g=4$ or $g=2$ in the scenarios described earlier), and $L_k$ and $D_k$ are the Lifshitz-Kosevich and Dingle factors describing the suppression of the oscillation amplitude by temperature and disorder broadening respectively, which are explicitly given by:

\be
L_k=\frac{ \frac{2\pi^2 k}{\beta \omega_c}}{\sinh\left( \frac{2\pi^2 k}{\beta \omega_c} \right)} , \  D_k = e^{-\pi k \Gamma/\omega_c }.
\ee

From the above free energy the oscillations of various equilibrium thermodynamic properties can be obtained. For example, the magnetization oscillations follow from:

\be\label{Mdef}
{\bf M}=-\frac{\partial \Omega_{osc}}{\partial {\bf B}}.
\ee

In the above expression ${\bf B}$ is the physical magnetic field. Therefore in order to compute the magnetization one needs to know what is the dependence of the parameters of the spinon Fermi surface state, including the emergent magnetic field ${\mathcal B} _ 0$, as a function of ${\bf B}$. We will use some phenomenological guide from the experiments that measure oscillations of thermal conductivity~\cite{OngQO} to predict the oscillations of magnetization. Let us assume that within some region of parameters in which the pseudoscalar spinon Fermi surface is realized, the main parameter changing with the physical field in-plane is the emergent magnetic field ${\mathcal B} _ 0$. For example when the physical field is along ${\bf b}$ axis (${\bf B} || {\bf b}$), let us assume that the emergent field grows linearly with the component of physical magnetic field along such axis denoted by $B_{\bf {\rm b}}$:

\be
{\mathcal B} _ 0 \approx \alpha B_{\bf {\rm b}}.
\ee

The above linear dependence is consistent with the experimental observation that the thermal conductivity has nearly equally spaced oscillations when plotted as a function of $1/B_{\bf {\rm b}}$~\cite{OngQO}. Then Eq.\eqref{Mdef} predicts an oscillation of the induced magnetization along the ${\bf b}$ axis given by:

\be
M^{\rm osc}_{\bf {\rm b}} \approx \alpha \frac{E_0}{2 \pi^2} \sum_ {k = 1}^{\infty} L_k D_k \frac{(-1)^{k+1}}{k} \sin \left(\frac{kS}{|{\mathcal B}_0|}\right).
\ee

Where we have made the usual approximation of only taking the derivatives with respect to ${\mathcal B} _ 0$ inside the argument of the cosine function in Eq.~\eqref{Omegaosc}, since the other contributions are subdominant when $|{\mathcal B} _ 0| \ll S$. To close this section, we wish to note that the above computation was performed imagining the minimal set of either 4 or 2 pockets related by $M$ and $C$ symmetries described earlier. If there are other pockets which are also small and elliptical but not related by symmetry to these, one can simply add the separate contributions.

\section{Discussion and outlook}
\label{discussion}

\subsection{On theories and models}

In this work we have introduced the notion of pseudoscalar U(1) spin liquids employing the fermionic parton representation of spins and a class of Gutzwiller-projected trial wavefunctions parametrized by spinon Slater determinants. This approach is a convenient tool, but it should not be viewed as the definition of the state of matter itself. Pseudoscalar U(1) spin liquids can more generally be regarded as states in which a microscopic symmetry acts as a particle-hole conjugation on the non-local emergent fractionalized spinon particles. Are there other theoretical approaches that might allow us to investigate these states? We would like to offer examples of some ideal models that suggest that perhaps pseudoscalar spinons are more common than previously recognized and perhaps some of these models could offer clues and more amenable playgrounds to better understand the microscopic underpinnings of the emergence of these states.

In fact, it is possible to show that the spinons that emerge in 1D from the standard Jordan-Wigner transformation that maps the 1D XXZ Heisenberg model at $J_z=0$ onto free fermions are in a sense pseudoscalar spinons. One can in particular show that for example the mirror symmetry that reverses the axis direction of the 1D chain, acts on the Jordan-Wigner fermion operators as a particle-hole conjugation. Although in this case there is not properly speaking a U(1) gauge structure, one can define an analogue of magnetic flux through the loop of the 1D chain when it is placed in periodic boundary conditions, and one can show that in fact this magnetic field is invariant under the aforementioned mirror, in contrast odd transformation for the magnetic field in the case of a mirror acting on an ordinary 1D electrons wire~\cite{unpublished}. Therefore the traditional XXZ Heisenberg model with Jordan-Wigner fermions can be regarded as a 1D toy version of a pseudoscalar spin liquid state with gapless spinons.

It is also possible to regard the charges in certain quantum spin ice as models as pseudoscalar spinons. In three dimensional quantum spin ice~\cite{HermeleICE} this can be rationalized by taking the charges defined by the ice rule to be the magnetic monopoles~\cite{Castelnovo}, so that their pseudoscalar nature follows from what is naturally expected from the transformation laws of ordinary magnetic fields. However in two dimensions the distinction becomes more non-trivial, because the electric and magnetic fields can be clearly distinguished, since the electric field is a two-component in-plane vector while the magnetic field is a single-component scalar. Therefore in two-dimensional quantum spin ice~\cite{Shannon} or the closely related two-dimensional quantum dimer model~\cite{RK,6vertex}, one can show that the spinon number defined as the charge associated with the ice rule, is indeed odd under mirror operations and thus certain U(1) spin liquid states emerging in these models could also be regarded as pseudoscalar spin liquids~\cite{unpublished}. 


We would also like to contrast the pseudoscalar spinon Fermi surface states described in this work with the ``composite exciton fermi liquid"  states introduced in Ref.~\cite{Debanjan}. Both of these states feature spinon particle-like and hole-like Fermi surfaces with equal sizes. The compensation of particle and hole pockets in both states is a consequence of having an even number of spinons per unit cell, regardless of point-group crystalline symmetries. However the composite exciton fermi liquid is ``ordinary" or ``scalar" in the sense that the emergent magnetic field has the same transformation laws expected for the usual physical magnetic field experienced by electrons. Therefore the pseudoscalar spinon Fermi surface states and the composite exciton fermi liquid are sharply distinct phases of matter with regard to symmetry that can be regarded as distinct symmetry enriched U(1) gapless spin liquids. In particular, the composite exciton fermi liquids, as introduced in Ref.~\cite{Debanjan}, would generically feature coexisting finite thermal Hall effect and quantum oscillations when the applied field is along the ${\bf b}$ axis of $\alpha$-RuCl$_3$, and therefore these are not natural candidates to explain the phenomenology reported in Ref.\cite{OngQO}.

\subsection{On experiments and materials}

We begin by commenting on another set of prominent experiments that has argued for the presence of a chiral spin liquid with gapless Majorana edge modes and half-quantized thermal Hall conductivity for fields along the ${\bf a}$ axis~\cite{Matsuda_2018B,Matsuda_2020,Takagi_2021}. At the moment the precise connection to the study of Ref.~\cite{OngQO} is not completely clear to us, since the latter did not find clear evidence of the quantization of the thermal Hall conductivity. However, the more recent study of Ref.~\cite{Takagi_2021}, provided evidence that such quantized state would survive to lower temperatures at fields along the ${\bf a}$ axis that are above $\sim11T$, which is somewhat above the regime in which Ref.~\cite{OngQO} focused on. It is important also to note that the quantized thermal Hall conductivity argued in Refs.~\cite{Matsuda_2018B,Matsuda_2020,Takagi_2021} is observed above the temperatures at which the quantum oscillations clearly set in Ref.~\cite{OngQO} for the region of fields below $\sim 11T$. Therefore, while as a matter of principle chiral spin liquids could descend from the parent pseudoscalar spin liquid as a result of the Landau quantization, the above observation is indicative that if the Majorana chiral spin liquid is present, it might be competing with the pseudoscalar spinon fermi surface state and not necessarily descend from it. We caution however that the previous statement implicitly relies on weak coupling intuition, since it imagines the cyclotron spacing energy scale that describes the spinon Landau level spectrum as a distinct energy scale from the spinon interactions, and implicitly assumes that the latter can be viewed as a perturbation in comparison to the former. But as we have emphasized before, the spinons are generically strongly coupled by the gauge fields, and there is no simple generic way to deform the problem into such regime of separation of spinon single-particle and interaction scales. We would therefore like to encourage future experiments to further investigate the relations and competitions between these fascinating states of matter.

We would also like to offer some suggestions also on possible future experiments, because we believe there is still much to be learned experimentally on the spin liquid states realized in $\alpha$-RuCl$_3$. One relatively immediate additional information that could be provided by thermal conductivity measurements are the full two components of the longitudinal conductivity. The current measurements reported only the longitudinal conductivity along the direction of the in-plane field, but there should be an independent value of the longitudinal conductivity perpendicular to the magnetic field (assuming that the field is along either the ${\bf a}$ or ${\bf b}$ axis). This extra component could offer clues on how anisotropic the spin liquid state is. Another important question would be to develop a more detailed global map of the evolution of the properties of the spin liquid states as the in-plane field is rotated. Torque magnetometry that detects magnetization oscillations could be a versatile probe allowing to get more details on the full in-plane field orientation dependence of the period and temperature dependence of the amplitude of oscillations. This could help clarify the precise relation among the spin liquids realized when the in-plane field is along the ${\bf a}$ and ${\bf b}$ axis and in-between. The recent study of Ref.~\cite{Takagi_2021} also indicates that it is important to clarify the behavior of the large in-plane field state realized above $\gtrsim 11T$ and its potential connection to the Majorana chiral spin liquid state. 

It would also be interesting to contemplate the possibility that other spinon Fermi surface candidate materials, that have displayed some phenomenology at odds with the traditional spinon Fermi surface scenario, might harbour pseudoscalar spinon Fermi surface states. One notable example are the organic materials that have been reported to not display quantum oscillations with applied perpendicular field~\cite{organics}, in spite of displaying other properties consistent with a spinon Fermi surface scenario~\cite{KanodaRMP}. It will be also interesting to examine the potential relevance of the pseudoscalar spinon fermi surface scenarios to the Kondo and Anderson periodic lattice problems~\cite{Coleman}, specially since several heavy-fermion compounds with even number of electrons per unit cell have been shown to display quantum oscillations or features consistent with spinon Fermi surface state~\cite{Li,Sebastian,MatsudaYbB12}. Furthermore, it will be interesting to investigate the potential relevance of the pseudoscalar U(1) spin liquids in the context of transition metal dichalcogenides that have displayed quantum oscillations coexisting with insulating behavior~\cite{Wu} and also some phenomenology consistent with spinon Fermi surface states~\cite{Lee2017,Lee2018,Matsuda2020}.

\begin{acknowledgments}
We thank Paul McClarty, Peter Czajka and Nai Phuan Ong for stimulating discussions.

\end{acknowledgments}

\end{document}